\documentstyle{article}
\begin{document}
\title{REFLECTION TIME AND THE GOOS-H\"{A}NCHEN EFFECT \\
FOR REFLECTION BY A SEMI-INFINITE RECTANGULAR BARRIER \\[.25in]}

\author{Edward R. Floyd \\
{\it 10 Jamaica Village Road, Coronado, California 92118-3208} \\
floyd@crash.cts.com \\[.125in]}

\date{Received 1 November 1996;\ \ revised 26 April 2000 \\[.125in]
Version 3 \\
Final version consistent with galley proofs \\
To be published in {\it Found. Phys. Lett.}}

\maketitle

\begin{abstract}

The reflection time, during which a particle is in the classically forbidden
region, is described by the trajectory representation for reflection by a 
semi-infinite rectangular barrier.  The Schr\"{o}dinger wave function has
microstates for such reflection.  The reflection time is a function of the
microstate.  For oblique reflection, the Goos-H\"{a}nchen displacement is also
a function of the microstate.  For a square well duct, we develop a proposed test
where consistent overdetermination of the trajectory by a redundant set of
observed constants of the motion would be beyond the Copenhagen interpretation.
\\[.125in]

\end{abstract}


\smallskip


\bigskip

\section{INTRODUCTION}

Tunneling times, reflection times and dwell times have been well studied for
barriers.  Regrettably, we still do not have universally accepted answers for
these times; even the concepts for these times remain in dispute 
[\ref{bib:rmp61}--\ref{bib:pla197}].  Part of the problem is that motion in the
classically forbidden region inside the barrier is a quantum process that is
confounded by the interference between matter reflecting and tunneling.

Herein, we investigate reflection of particles with sub-barrier energy by a 
semi-infinite rectangular barrier.  This simplifies the problem as all particles
are reflected only. We use a trajectory representation 
[\ref{bib:prd26}--\ref{bib:fpl}] to derive the reflection time.  We compare our
results with those extrapolated from other investigations of tunneling.  We first
study reflection time for normal incidence to show that the trajectory method is
valid as it produces reflection times that are consistent with other results. 
We next study the Goos-H\"{a}nchen displacement for oblique incidence.  We then
consider square wells and ducts. The Goos-H\"{a}nchen displacements for a square
well duct and other constants of the motion present a proposed, strong test for
resolving the ``underdetermination" issue [\ref{bib:jtc}] whether the trajectory
representation and the Copenhagen interpretations may be distinguished.  This
proposed test hinges on whether an overdetermined set of constants of the motion
is consistent in observation with the theoretical redundance of this set.  

This investigation is an application of the trajectory representation.  We begin
with the generalized Hamilton-Jacobi equation for the trajectory and its
solution, a generalized Hamilton's characteristic function, which is the
generator of the motion [\ref{bib:prd34},\ref{bib:fpl}].  Application of the
trajectory representation to a quantum problem follows a recipe for processing
the generalized Hamilton's characteristic function of quantum mechanics which in
practice is similar to the recipe for processing the classical Hamilton's
characteristic function for the analogous problem in classical mechanics.  This
recipe is standard; in both cases, the Hamilton-Jacobi transformation equations
for constant coordinates (often called Jacobi's theorem) are the equations of
motion.  Those unfamiliar with the trajectory representation who are interested
in the theoretical foundations of the trajectory representation can find a
progressive treatment in references \ref{bib:prd26} through \ref{bib:fpl}.  The
trajectory representations differs with Bohmian mechanics [\ref{bib:pr85}] even
though both spring from essentially the same generalized Hamilton-Jacobi
equation.  These differences have been discussed elsewhere
[\ref{bib:prd26},\ref{bib:prd29},\ref{bib:pe7}].  In this exposition, we shall
note significant differences as they occur.
    
We have chosen to examine the semi-infinite rectangular barrier because its
Schr\"{o}dinger representation is familiar.  Separation of variables applies. 
Furthermore, exact solutions exist in closed form in terms of elementary
functions for both the trajectory and Schr\"{o}dinger representations.

Microstates of the Schr\"{o}dinger wave function exist for reflection by a 
semi-infinite barrier because this barrier induces a nodal singularity in the
trajectory representation that is associated with a non-isolated zero in the
Schr\"{o}dinger wave function [\ref{bib:fpl}].  We show that the reflection time
and the Goos-H\"{a}nchen displacement are deterministic.  They depend upon the
particular microstate of the reflecting particle.   A reflection time or a 
Goos-H\"{a}nchen displacement that is a function of the particular microstate is
a manifestation that the Schr\"{o}dinger wave function is not in general an
exhaustive description of nonrelativistic phenomenon.  

We include in this investigation a study of oblique incidence where there is a
Goos-H\"{a}nchen displacement between the incident trajectory's entry point and
the reflected trajectory's exit point in the barrier [\ref{bib:ap1}].  We show
that the Goos-H\"{a}nchen displacement is dependent upon the particular
microstate.  The Goos-H\"{a}nchen displacement is presented as a candidate to
sidestep the conceptual difficulties in tunneling with ``clocks"
[\ref{bib:rmp66}] and the absence of a time operator in the operator formulation
of nonrelativistic quantum mechanics [\ref{bib:prl74}].  Therefore, the 
Goos-H\"{a}nchen displacement offers a better test for resolving the
underdetermination than reflection time.

Our investigation of oblique incidence also shows that the trajectory
representation is not a hydrodynamic representation, such as Bohmian mechanics,
for the trajectories are not stream lines as they become imbedded in the surface
of constant Hamilton's characteristic function at the turning point infinitely
deep inside the barrier.  This allows velocities to increase without bound along
the trajectory at the turning point.  This also induces the turning points to be
cusps.

The reflection time is finite for either normal or oblique incidence while the
distance traversed by the trajectory is infinite.  Therefore, the velocity along
the trajectory must be infinite over a measurable duration.  

We also examine a square well and a square well duct which are extensions of the
semi-infinite barrier problem.  We report that the reflection time as well as the
period of libration are determined by the particular microstate.  We show that
these times are functions of various constant of the motion.

The trajectory representation is causal and not based upon chance
[\ref{bib:prd26}--\ref{bib:fpl}].  The trajectory representation has shown, by
certain tunneling through a finite barrier by a particle with sub-barrier energy,
that the Born postulate of the Copenhagen interpretation, which attributes a
probability amplitude to the Schr\"{o}dinger wave function, is unnecessary
[\ref{bib:afb20}].  Any solitary trajectory or microstate is sufficient by itself
to specify the Schr\"{o}dinger wave function [\ref{bib:prd34}].  The set of
initial conditions that are necessary and sufficient to specify the particular
trajectory or microstate are known [\ref{bib:prd29}].  There is no need to invoke
an ensemble of trajectories to get the Schr\"{o}dinger wave function or to
describe quantum phenomena.  The trajectory representation renders predictions
for an individual particle viz-a-viz the probability predictions of the
Copenhagen representation for an ensemble of particles [\ref{bib:afb20}].  While
both the trajectory representation and Bohmian mechanics are causal, Bohmian
mechanics is still a stochastic theory [\ref{bib:pr172}] where an ensemble of
trajectories are needed to describe quantum phenomena.

Without a stochastic requirement, the trajectory representation does not need a
wave packet to describe or localize a particle.  This simplifies our
investigation by allowing us to use a monochromatic particle.  This also
generalizes our findings by making them independent of the shape of a wave
packet.  

In this investigation, we apply the Schr\"{o}dinger representation without the
Copenhagen interpretation which assigns a probability amplitude to the
Schr\"{o}dinger wave function.  Lest we forget, Schr\"{o}dinger opposed the
Copenhagen interpretation of his wave function.

For completeness, scattering of particles with super-barrier energy from a 
semi-infinite barrier has been reported elsewhere [\ref{bib:pe7}].

In Section 2, we develop the equations of motion for normal incidence, determine
the reflection time, and show that it is dependent upon the particular
microstate.  We develop the corresponding Schr\"{o}dinger representation as far
as it goes to show that it does not discern the dependence due to microstates. 
In Section 3, we develop the trajectory equations for oblique incidence and
determine the Goos-H\"{a}nchen displacement.  We also demonstrate that the
trajectories are generally not wave normals.  In Section 4, we develop the
reflection time and the libration period for the square well.  We show that
reflection time and libration period are described by constants of the motion
that are beyond the Copenhagen interpretation.  In Section 5, we investigate
square well ducts.  We develop the Goos-H\"{a}nchen displacement and libration
displacement as constants of the motion for a trajectory in the square well duct. 
We show that overdetermining the microstate (trajectory) in a duct renders a
strong test for resolving underdetermination.

\section{NORMAL INCIDENCE}

\paragraph{Equation of Motion:} Let us initially consider the reflection of a
particle that is normally incident to a semi-infinite rectangular barrier whose
potential is given by

\[
V(x) = \left\{ \begin{array}{cc}
               0, & \  \  x<0 \\ [.1in]
               U, & \  \  x\geq 0
               \end{array}
       \right.
\label{eq:ve}
\]

\noindent where $U$ is finite positive.  For normal incidence, the particle will
have motion only in the $x$-direction.  We choose a sub-barrier energy given by
$E_x=(\hbar k_x)^2/(2m)<U$ where $E_x$ is the energy for an incident particle,
$k_x$ is the wavenumber in the $x$-direction, $m$ is mass of the particle, and
$\hbar=h/(2\pi)$ where in turn $h$ is Planck's constant.  For normal incidence,
we make it explicit by using $E_x$ that only the $x$-dimension contributes to
energy.  

The trajectory representation is based upon a generalized Hamilton-Jacobi
formulation.  The time-independent generalized Hamilton-Jacobi equation for
quantum mechanics is a phenomenological equation that is given for 
one-dimensional motion in the $x$-direction with normal incidence by
[\ref{bib:prd29}--\ref{bib:fpl}] 

\begin{equation}
\frac{(\partial W/\partial x)^2}{2m}+V-E_x=-\frac{\hbar ^2}{4m}\langle W;x\rangle
\label{eq:hje}
\end{equation}

\noindent where $\langle W;x\rangle $ is the Schwarzian derivative of $W$ with
respect to $x$.  The Schwarzian derivative is given by

\[
\langle W;x\rangle = \left[\frac{\partial ^3W/\partial x^3}{\partial W/\partial
x}-\frac{3}{2}\left(\frac{\partial ^2W/\partial x^2}{\partial W/\partial
x}\right)^2 \right].
\]

\noindent In Eq.\ (\ref{eq:hje}), $W$ is Hamilton's characteristic function,
$\partial W/\partial x$ is the momentum conjugate to $x$.   The left side of Eq.\
(\ref{eq:hje}) manifests the classical Hamilton-Jacobi equation while the
Schwarzian derivative on the right side manifests the higher order quantum
effects.  We explicitly note that $W$ and $\partial W/\partial x$ are real even
in the classically forbidden region inside the barrier $(x\geq 0)$.  The general
solution for $\partial W/\partial x$ is given by [\ref{bib:prd34},\ref{bib:fpl}] 

\begin{equation}
\partial W/\partial x = \pm (2m)^{1/2}(a\phi ^2+b\theta ^2+c\phi \theta )^{-1}
\label{eq:cme}
\end{equation}

\noindent where $(a,b,c)$ is a set of real coefficients such that  $a,b > 0$, and 
$(\phi,\theta)$ is a set of normalized independent solutions of the associated
time-independent one-dimensional Schr\"{o}dinger equation,

\[
\frac{-\hbar ^2}{2m} \frac{d^2\psi }{dx^2} + (V-E_x)\psi = 0. 
\]

\noindent The independent solutions $(\phi,\theta)$ are normalized so that their
Wronskian, ${\cal W}(\phi,\theta) = \phi \, d\theta /dx - d\phi /dx\, \theta $,
is scaled to give  ${\cal W}^2(\phi,\theta) = 2m/[\hbar ^2(ab-c^2/4)] > 0$.  (
The nonlinearity of the generalized Hamilton-Jacobi equation induces this
normalization upon {\cal W}.)\ \, This ensures that $(a\phi ^2 + b\theta ^2 +
c\phi \theta) > 0$.  Also, the conjugate momentum is not the mechanical momentum,
i.e., $\partial W/\partial x \ne m\dot{x}$.  We note for completeness that a
particular set $(\phi,\theta)$ of independent solutions of the Schr\"{o}dinger
equation may be chosen by the superposition principle so that the coefficient $c$
is zero.  

The motion in phase space is specified by Eq.\ (\ref{eq:cme}).  This phase-space
trajectory is a function of the set of coefficients $(a,b,c)$.  The $\pm$ sign
in Eq.\ (\ref{eq:cme}) designates that the motion may be in either $x$-direction. 
  
The corresponding solution for the generalized Hamilton's characteristic
function, $W$, is given by

\begin{equation}
W=\hbar \arctan \left(\frac{b(\theta /\phi ) + c/2}{(ab-c^2/4)^{1/2}}\right)+K
\label{eq:hcfe}
\end{equation}

\noindent where $K$ is an integration constant which we may set to zero herein.

Hamilton's characteristic function is a generator of motion.  The equation of
motion in the domain $[x,t]$ is given by the Hamilton-Jacobi transformation
equation for a constant coordinate.  The procedure simplifies for coordinates
whose conjugate momenta are separation constants. For stationarity, $E_x$ is a
separation constant for time for a trajectory with normal incidence.  Thus, the
equation of motion for the trajectory time, $t$, relative to its constant
coordinate $\tau$, is given as a function of $x$ by 

\begin{equation}
t-\tau= \partial W/\partial E_x
\label{eq:eom}
\end{equation}

\noindent where the trajectory for a given energy, $E_x$, is a function of a set
of coefficients $(a,b,c)$ and $\tau $ specifies the epoch.  The equation of
motion for the trajectory, Eq.\ (\ref{eq:eom}), and the Bohmian equation of
motion differ.  Bohmian mechanics asserts that $\partial W/\partial x$ in Eq.\
(\ref{eq:hje}) would be the mechanical momentum, $m\dot{x}$ and the subsequent
integration of Eq.\ (\ref{eq:hje}) would become Bohm's equation of motion
[\ref{bib:pr85}].  

\paragraph{Microstates:}  A particle with normal incidence and with sub-barrier
energy, $E_x$, has its turning point at $x=\infty$ where the Schr\"{o}dinger wave
function goes to a non-isolated zero [\ref{bib:prd34},\ref{bib:fpl}].  In the
generalized Hamilton-Jacobi representation, the barrier induces a nodal
singularity at $x=\infty $ in $\partial W/\partial x$ [\ref{bib:fpl}].  From Eq.\
(\ref{eq:cme}), $\partial W/\partial x \rightarrow 0$ as $x \rightarrow \infty
$, regardless of the values of the coefficients $a,\ b$ and $c$, because at least
one independent solution of the set $(\phi,\theta)$, in the classically forbidden
region inside the barrier, must increase without bound as $x \rightarrow \infty$. 
Each trajectory, which is specified by the particular values of the coefficients
$a$, $b$ and $c$, is a particular microstate of the Schr\"{o}dinger wave function
[\ref{bib:prd34},\ref{bib:fpl}].  Each microstate, by itself, is sufficient to 
specify the quantum results of a single event; there is no need to invoke an
ensemble of microstates to describe quantum phenomenon [\ref{bib:pla214}].

\paragraph{Reflection:}  A set of independent solutions $(\phi ,\theta )$ for our
given semi-infinite rectangular barrier can be chosen as 

\begin{equation}
\phi  =  \left(\frac{2m}{\hbar ^2k_x^2(ab-c^2/4)}\right)^{1/4} \cdot \left\{
\begin{array}{lr}
                   \cos [k_xx + \arctan(\kappa /k_x)], &  x<0 \\  [.08 in]
              \frac{{\displaystyle \exp (-\kappa x)}}{{\displaystyle [1+(\kappa
/k_x)^2]^{1/2}}}, & x \geq 0.
             \end{array}
             \right.
\label{eq:phi}
\end{equation}

\noindent and

\begin{equation}
\theta =  \left(\frac{2m}{\hbar ^2k^2(ab-c^2/4)}\right)^{1/4} \cdot \left\{
\begin{array}{lr}
\sin [k_xx+\arctan (\kappa /k_x)],  & x<0 \\ [.08 in]
\frac{{\displaystyle \left( \frac{\kappa }{k_x}+\frac{k_x}{\kappa }\right) \exp
(\kappa x) + \left( \frac{\kappa }{k_x}-\frac{k_x}{\kappa }\right) \exp (-\kappa
x)}}{{\displaystyle 2[1+(\kappa /k_x)^2]^{1/2}}}, & x \geq 0 \\
\end{array}
\right.
\label{eq:theta}
\end{equation}

\noindent where $\kappa = [2m(U-E_x)]^{1/2}/\hbar $.  The corresponding Wronskian obeys
${\cal W}^2(\phi ,\theta ) = 2m/[\hbar ^2(ab-c^2/4)] \geq 0$ as expected.

The conjugate momentum is given by Eq.\ (\ref{eq:cme}) as 

\[
\partial W(E_x,a,b,c,x)/\partial x=\pm (2m)^{1/2}[a\phi ^2(E_x,x)+b\theta
^2(E_x,x)+c\phi (E_x,x)\theta (E_x,x)]^{-1}
\]

\noindent where $\phi $ and $\theta $ are respectively given by Eqs.\
(\ref{eq:phi}) and (\ref{eq:theta}) and where the dependence of the conjugate
momentum upon energy, $E_x$, and the set of coefficients $(a,b,c)$ is made
explicit.  By Eqs.\ (\ref{eq:cme}), (\ref{eq:phi}) and (\ref{eq:theta}),
$\partial W/\partial x$ and $\partial ^2W/\partial x^2$ are continuous across the
barrier interface at $x=0$.  As expected, we see from Eq.\ (\ref{eq:theta}) that
the $\exp (\kappa x)$ term in $\theta $ will induce a nodal singularity in
$\partial W/\partial x$, Eq.\ (\ref{eq:cme}), for any allowed values of the
coefficients $a,\ b$ and $c$ at the trajectory's turning point at $x=\infty$.

We define the reflection time, $t_R$, herein as the duration that a particle
spends in the classically forbidden region inside the barrier, which is given by
the round trip time for the particle between the barrier surface at $x=0$ and the
turning point at $x=\infty $.  From the equation of motion, Eq.\ (\ref{eq:eom}),
and Eqs.\ (\ref{eq:hcfe}), (\ref{eq:phi}) and (\ref{eq:theta}), the reflection
time, $t_R=2[t(\infty )-t(0)]$, is given by 

\begin{equation}
t_R=2 \frac{(ab-c^2/4)^{1/2} [1+(\kappa /k_x)^2]} {a + c (\kappa /k_x) + b(\kappa
/k_x)^2} \frac{m}{\hbar \kappa k_x}.
\label{eq:rte}
\end{equation}

\noindent The reflection time, $t_R$, is dependent upon the particular trajectory 
or microstate as specified by the coefficients $a$, $b$ and $c$.  We find that
$t_R$ is inversely proportional to $\kappa $ consistent with Barton
[\ref{bib:ap166}] who found that tunneling time, $t_T$, decreased with increasing
$\kappa $ for wave packets tunneling through inverted oscillator barriers.

The reflection time is finite because $\kappa $ and $k_x$ are real and ${\cal
W}^2 >0$.  But the trajectory traverses an infinite distance between the
interface at $x=0$ and the turning point at $x=\infty $ in a finite duration of
time.  Hence, the velocity along the trajectory must be infinite for a measurable
duration of time.  Note that this infinite velocity exists even though
$\lim_{x\rightarrow \infty } \partial W/\partial x \rightarrow 0$ as already
noted.  (We discuss later an analogous effect in higher dimensions, cf.\ \S \ 3,
\P \  Trajectory Directions and Wave Normals). 

As already noted, all trajectories, regardless of the particular values of the
coefficients, $a,\ b$ and $c$ have their turning point at $x=\infty $ because the
solution, $\partial W/\partial x$, to the generalized Hamilton-Jacobi equation,
Eq.\ (\ref{eq:hje}), has a nodal point there.  The set of coefficients ($a,b,c$)
can be specified the set of initial conditions ($x_o,\dot{x}_o,\ddot{x}_o$) for
a particle at some finite $x_o$ [\ref{bib:prd34}].  Even if we could not
prescribe the set of initial conditions
($x_o,\dot{x}_o,\ddot{x}_o$) (due perhaps to limits of practicality, but not to
any limits of principle for the trajectory representation), then we would have
to determine the distribution of the coefficients $a,\ b$ and $c$.   Any
distribution of the reflection time, $t_R$, over an ensemble of different
microstates is due to the particular distribution of the coefficients $a,\ b$ and
$c$ and not due to any adduced distribution of early turning points at finite
depths as hypothesized by quantum interpretations that attribute a probability
amplitude to $\psi $.  In general, these distributions will not be the same.

Let us now examine $t_R$ for a particular case.  Let us assume that $a=b$ and
$c=0$. In such case, the reflection time, Eq.\ (\ref{eq:rte}), becomes

\[
t_R = 2m/(\hbar \kappa k_x) = \hbar /[E_x(U-E_x)]^{1/2}.
\]

\noindent This finding for monochromatic propagation is consistent with findings
for tunneling by wave packets.  For rectangular barriers, Hauge and Stovneng
[\ref{bib:rmp61}] and Olkhovsky and Recami [\ref{bib:pr214}] have shown for phase
times and for Larmor times respectfully that $t_R=t_T$ where $t_T$ is tunneling
time.  For arbitrarily large thick barriers, Hartman [\ref{bib:jap33}] has shown
that $t_T \approx 2m/(\hbar \kappa k_x)$ for spatial Gaussian wave packets, and
Fletcher [\ref{bib:jpc18}] has shown that $t_T = \hbar /[E_x(U-E_x)]^{1/2}$ for
wave packets with temporal exponential leading and trailing tails.

As the trajectory representation is consistent with other work, we have
confidence in applying it to less studied situations in \S\ 3--5.    

\paragraph{Schr\"{o}dinger Representation:} We now investigate the corresponding
Schr\"{o}dinger representation for monochromatic reflection by a semi-infinite
rectangular barrier.  As all particles are reflected, the Schr\"{o}dinger wave
function for the $x$-component can be represented by a real function within an
arbitrary phase factor.  The Schr\"{o}dinger wave function, $\psi $, can be
represented in trigonometric form as [\ref{bib:prd34},\ref{bib:fpl}]

\begin{eqnarray}
\psi & = & \frac{(2m)^{1/4} \cos(W/\hbar )}{[a-c^2/(4b)]^{1/2}(\partial
W/\partial x)^{1/2}} \nonumber \\
& = & \frac{(a\phi ^2+b\theta ^2+c\phi \theta )^{1/2}}{[a-c^2/(4b)]^{1/2}} \cos
\left[\arctan \left(\frac{b(\theta /\phi ) + c/2}{(ab-c^2/4)^{1/2}}\right)\right]
= \phi . 
\label{eq:tre}
\end{eqnarray}

\noindent where $\psi $ and $d\psi /dx$ are continuous across the barrier
interface at $x=0$ as described by Eq.\ (\ref{eq:phi}).  While $\psi =\phi $
where $\phi $ is independent of the coefficients $a$, $b$ and $c$, the generator
of motion, $W$, for each particular trajectory (a microstate of $\psi $) is a
function of the coefficients $a$, $b$ and $c$ is given by Eq.\ (\ref{eq:hcfe}). 
Hence, the Schr\"{o}dinger wave function, $\psi $, is not an exhaustive
description of reflection of a particle with sub-barrier energy by a 
semi-infinite rectangular barrier.

The intermediate steps in Eq.\ (\ref{eq:tre}) systematically inject the
coefficients ($a,b,c$) into the amplitude and phase of $\psi $ so that $\psi $ 
still remains independent of these coefficients.  Nevertheless, ($a,b,c$)
determine the microstate of $\psi $.  Each microstate of $\psi $ has its
distinguishing amplitude and phase modulation determined by the coefficients
($a,b,c$).   We note that the Copenhagen interpretation asserts that $\psi $
would be the exhaustive description of nonrelativistic quantum phenomenon. 
Therefore, the Copenhagen school would declare the intermediate steps in Eq.
(\ref{eq:tre}) to be extraphysical.  Again, we apply the Schr\"{o}dinger
representation herein without the Copenhagen interpretation.

We can represent $\psi $ in an exponential format as an alternative to Eq.\
(\ref{eq:tre}) by

\[
\psi = \frac{(a\phi ^2+b\theta ^2+c\phi \theta )^{1/2}}{2[a-c^2/(4b)]^{1/2}}
\left\{ \exp \left[i \arctan \left(\frac{b(\theta /\phi ) + c/2}{(ab-
c^2/4)^{1/2}}\right)\right] + \exp \left[-i \arctan \left(\frac{b(\theta /\phi
) + c/2}{(ab-c^2/4)^{1/2}}\right)\right]\right\}
\]

\noindent where the first and second exponential terms represent respectfully the
incident and reflected waves.  Although $\psi $ in its trigonometric
representation, Eq.\ (\ref{eq:tre}), is independent of the coefficients $a$, $b$
and $c$, this is not the case for exponential case when the incident and
reflected waves are considered individually.  Neither the incident wave nor the
reflected wave, when considered separately, are independent of these coefficients
for [\ref{bib:fpl},\ref{bib:pe7}]

\begin{equation}
\frac{(a\phi ^2+b\theta ^2+c\phi \theta )^{1/2}}{2[a-c^2/(4b)]^{1/2}}  \exp
\left[\pm i \arctan \left(\frac{b(\theta /\phi ) + c/2}{(ab-
c^2/4)^{1/2}}\right)\right] =  \frac{[1\pm ic/(4ab-c^2)^{1/2}]}{2}\phi \pm
\frac{ib}{2(ab-c^2/4)^{1/2}}\theta .
\label{eq:pmere}
\end{equation}

When $a=b$ and $c=0$ as in the case studied by Hauge and Stovneng
[\ref{bib:rmp61}] and Olkhovsky and Recami [\ref{bib:pr214}], then in the
classically allowed region, $x<0$, the generator of motion simplifies to

\[
W(E_x,a,b,c,x)\Big|_{a=b,c=0,x<0}=\hbar \arctan \{ \tan [k_xx + \arctan (\kappa
/k_x)]\} .
\]

\noindent This generator of the motion by Eq.\ (\ref{eq:tre}) is, within a
constant phase factor, consistent with an incident plane wave given by

\[
\frac{(2m)^{1/4}(k_x-i\kappa )}{2[\hbar k_x(k_x^2+\kappa ^2)]^{1/2}} \exp(ik_xx),
\]

\noindent and a reflected plane wave given by

\[
\frac{(2m)^{1/4}(k_x+i\kappa )}{2[\hbar k_x(k_x^2+\kappa ^2)]^{1/2}} \exp(-
ik_xx).
\]

\noindent This is the particular microstate, where $a=b$, and $c=0$, that
contemporary physics tacitly assumes when working in the Schr\"{o}dinger
representation [\ref{bib:jap33},\ref{bib:jpc18}].  This particular microstate
manifests rectilinear propagation for both the incident and reflected unmodulated
plane waves.  The incident and reflected waves for a more general microstate,
where $a\neq b$ or $c\neq 0$, have compound modulation in both amplitude and
wavenumber by Eq.\ (\ref{eq:pmere}) [\ref{bib:pe7},\ref{bib:afb20}].  These waves
with compound modulation are still solutions to the Schr\"{o}dinger equation
[\ref{bib:pe7}].

\paragraph{Underdetermination:}  As the Copenhagen interpretation does not
entertain microstates of $\psi $, the Copenhagen predictions for reflection time,
$t_R$, will differ with the trajectory predictions of reflection time, Eq.\
(\ref{eq:rte}), which are dependent upon the microstate for the trajectory $t_R$
is a function of the coefficients ($a,b,c$).

\paragraph{Bohmian Mechanics:}  As the $x$-component of the wave function is real
within a constant phase factor, Bohmian mechanics asserts that $W$ would be
independent of $x$ [\ref{bib:pr85}].  Therefore, Bohmian mechanics concludes that
the particle would stand still at its initial position, inside or outside the
barrier or at the barrier interface, and particle reflection would not occur.  

\section{OBLIQUE INCIDENCE}  

\paragraph{Trajectory Equation:}  Let us now examine the more general case of
oblique incidence.   We can always choose our cartesian coordinate system so that
the trajectory lies in the $x,y$-plane.  The generalized Hamilton-Jacobi equation
and the Schr\"{o}dinger equation are separable in cartesian coordinates.  In a
Hamilton-Jacobi representation, the cartesian coordinate $y$ is cyclic with a
transformed constant $y$-momentum given by $\hbar k_y$.  (This prescribes that
$y$-motion is constant or the $y$-component in the Schr\"{o}dinger representation
is an unmodulated wave.)  The cartesian coordinate $z$ is also cyclic but the
transformed constant $z$-momentum is zero for our choice of orientation for the
coordinate system.  The generalized Hamilton-Jacobi equation for oblique
incidence is given by

\[
\frac{(\partial W/\partial x)^2+(\partial W/\partial y)^2}{2m}+V-E=-\frac{\hbar
^2}{4m}\langle W;x\rangle
 \label{eq:ohje}
\]

\noindent where $E$ is the energy for a particle with oblique incidence.  As 
$y$-momentum is constant, any contribution due to a latent $\langle W;y\rangle
$ is zero.  The solution for the generator of the motion is given by 

\begin{equation}
W=\hbar \arctan \left(\frac{b(\theta /\phi ) + c/2}{(ab-c^2/4)^{1/2}}\right) +
\hbar k_yy.
\label{eq:ocfe}
\end{equation}

\noindent In the above, $\phi $ and $\theta $ are still given by Eqs.\
(\ref{eq:phi}) and (\ref{eq:theta}) respectfully, but where now 

\[
k_x = \left( \frac {2mE}{\hbar ^2} - k_y^2 \right)^{1/2}
\]

\noindent and

\[
\kappa = \left( \frac{2m(U-E)}{\hbar ^2} + k_y^2 \right)^{1/2}.
\]

\noindent where $E$ is constrained so that $k_x$ and $\kappa $ are real. 
Equation (\ref{eq:cme}) is still valid for $\partial W/\partial x$. 
  
The trajectory equation, which describes the progress of the trajectory along the
cyclic coordinate $y$ as a function of $x$, is the equation of motion produced
by the Hamilton-Jacobi transformation equation for the constant (reference)
coordinate $y_o$ given by  

\begin{equation}
y-y_o=-\partial W/\partial (\hbar k_y).  
\label{eq:te}
\end{equation}

\paragraph{Goos-H\"{a}nchen Displacement:}  The lateral displacement, $\Delta y$,
between where the trajectory enters and exits the barrier is known as the 
Goos-H\"{a}nchen displacement [\ref{bib:ap1}].  By the trajectory equation, Eq.\
(\ref{eq:te}), the Goos-H\"{a}nchen displacement, $\Delta y = 2[y(\infty) -
y(0)]$, is given by   

\[
\Delta y = 2 \frac{(ab-c^2/4)^{1/2} [1+(\kappa /k_x)^2]}{a + c(\kappa /k_x) +
b(\kappa /k_x)^2} \frac{k_y}{\kappa k_x}.
\] 

\noindent Hence, the Goos-H\"{a}nchen displacement, for a given energy E and a
given $k_y$, is a function of the coefficients $a$, $b$ and $c$.  Analogous to
the time of reflection, $\Delta y$ is inversely proportional to $\kappa$ and
$k_x$.  The Goos-H\"{a}nchen effect gives us an alternative to reflection time
for describing the quantum effects of reflection.

\paragraph{Trajectory Directions and Wave Normals:}  We already know that the
turning point for the trajectory is at $x=\infty$ inside the semi-infinite
barrier.  We now investigate the behavior of the trajectory near the turning
point by studying Eq.\ (\ref{eq:te}) there.  As the trajectory approaches the
turning point, we have that $\partial W/\partial x$ goes to zero as 

\begin{equation}
\lim_{x\rightarrow \infty} \frac{\partial W}{\partial (\hbar k_y)}
\longrightarrow 4(ab-c^2/4)^{1/2} \frac{k_y}{k_x}\frac{x \exp(-2\kappa
x)}{1+(\kappa /k_x)^2}.
\label{eq:tde}
\end{equation}   

\noindent Thus, as $x\rightarrow \infty $, the trajectory becomes asymptotic to
the $x$-direction regardless of the values of the coefficients $a$, $b$ and $c$. 
Nevertheless, how $\partial W/\partial x$ goes to zero is still a function of the
coefficients $a$, $b$ and $c$.  Hence, the turning point at $x=\infty $ for the
trajectory is a symmetric cusp in the $x,y$-plane for all allowed values of $a$,
$b$ and $c$.  This is just another manifestation that $\partial W/\partial x$ has
a nodal singularity for a particle with sub-barrier energy reflecting from a 
semi-infinite barrier.  A symmetric cusp for a turning point at infinity, which
is formed by its two branches with mirror symmetry of each other that approach
a common asymptote, is consistent with traversing an infinite displacement in $x$
while transversing only a finite distance $y$.  It is also consistent with a
Schr\"{o}dinger wave function that decreases exponentially in the classically
forbidden region.

Meanwhile, the wave normal is given by $\nabla W$.   From Eq.\ (\ref{eq:ocfe}),
the wave normal is

\begin{equation}
\nabla W = (2m)^{1/2}(a\phi ^2+b\theta ^2+c\phi \theta )^{-1}{\bf \hat{\imath}}
+ \hbar k_y{\bf \hat{\jmath}}.
\label{eq:wne}
\end{equation}

\noindent At the turning point, we know that $\lim_{x\rightarrow \infty } (a\phi
^2+b\theta ^2+c\phi \theta )^{-1} = 0$ because at least one solution of the set
of independent solutions $(\phi ,\theta )$ must increase without bound at the
turning point at $x=\infty$.  Thus, we conclude from Eqs.\ (\ref{eq:tde}) and
(\ref{eq:wne}) that the trajectory's direction at the turning point is embedded
in a surface of constant $W$ because the wave normal is orthogonal to the
trajectory.  This is a manifestation that the trajectories are not stream lines
of a hydrodynamic representation of quantum mechanics.  We also conclude that,
when the trajectory direction is embedded in a surface of constant W, its
velocity increases without bound (cf. \S \ 2, \P \  Reflection).  

\paragraph{Underdetermination:}  The Goos-H\"{a}nchen displacement, $\Delta y$,
renders a better case than $t_R$ to resolve the underdetermination issue because
the Schr\"{o}dinger representation has an operator for $y$ but does not have one
for time $t$.  Thus, both the trajectory representation and the Copenhagen have
the means to determine $\Delta y$.  Again as the Copenhagen interpretation does
not entertain microstates of $\psi $, the Copenhagen predictions for 
Goos-H\"{a}nchen displacement, $\Delta y$, will differ with the trajectory
predictions of Goos-H\"{a}nchen displacement, Eq.\ (\ref{eq:rte}), which are
dependent upon the microstate for the trajectory $\Delta y$ is a function of the
set of coefficients ($a,b,c$).

\paragraph{Bohmian Mechanics:}  As the $x$-component of the wave function is real
within a constant phase factor, Bohmian mechanics asserts that $W$ would be
independent of $x$ [\ref{bib:pr85}].  Therefore, Bohmian mechanics concludes that
the particle would travel in the $y$-direction only and would never reflect off
the barrier.

\section{SQUARE WELLS}  

Let us now investigate reflection inside square wells as an extension of
reflection from semi-infinite rectangular barriers. The potential for a square
well may be given by 

\begin{equation}
V(x) = \left\{ \begin{array}{cc}
               U, & \  \  |x|>q \\ [.08 in]
               0, &  \  \  |x|\leq q
               \end{array}
       \right.
\label{eq:swv}
\end{equation}

\noindent where $q$ is finite positive and where $E_x,\ U,\ k_x$ and $\kappa $
maintain their previous definitions.  A finite square well always has at least
one symmetric bound state.  While our results, Eqs.\ (\ref{eq:rtesw}) and
(\ref{eq:lte}), are valid for both symmetric and antisymmetric bound states, we
discuss only the symmetric bound states to shorten the presentation.  We could
have arbitrarily chosen to present only the antisymmetric (or odd) bound states
just as easily except that not all square wells have sufficient size, that is
$(2mU)^{1/2}q/\hbar > \pi /2$, to ensure that an antisymmetric bound state
exists.  For bound states, $E_x<U$. Energy is quantized for symmetric bound
states by the familiar transcendental equation $\tan (k_xq) = \kappa /k_x$, which
can be established from either the quantization of the action variable, $J$, for
symmetric states by [\ref{bib:pla214}] 

\[
J = \oint \partial W/\partial x\, dx = (2n+1)h,\mbox{\ \ } n=0,1,2,\cdots \mbox{
and }n\pi <(2mU)^{1/2}q/\hbar 
\]

\noindent or the quantization energy, $E_x$, so that $\psi|_{x=\pm \infty}=0$. 
In either case, the quantization is independent of the coefficients $a,\ b$ and
$c$ [\ref{bib:prd34}].  The set of independent solutions $(\phi ,\theta )$ for
this square well is chosen such that $\phi $ represents the symmetric bound state
given by

\begin{equation}
\phi  =  \left(\frac{2m}{\hbar ^2k_x^2(ab-c^2/4)}\right)^{1/4} \cdot \left\{
\begin{array}{lr}
                   \cos (k_xq) \exp [-\kappa (x-q)], & x>q \\ [.08 in]
                   \cos (k_xx), & -q \leq x \leq q \\  [.08 in]
                   \cos (k_xq) \exp [\kappa (x+q)] & x<-q.
                   \end{array}
             \right.
\label{eq:phisw}
\end{equation}

\noindent The other solution, $\theta $, is unbound and is not unique as any
amount of $\phi $ may be added to it.  While $\phi $ is symmetric for the
symmetric bound state, the corresponding $\theta $ that we have chosen is
antisymmetric. We present this unbound solution as

\begin{equation}
\theta =  \left(\frac{2m}{\hbar ^2k_x^2(ab-c^2/4)}\right)^{1/4} \cdot \left\{
\begin{array}{lr}
\displaystyle{\frac{\exp [\kappa (x-q)] - \cos (2k_xq) \exp [-\kappa
(x+a)]}{2\sin (k_xq)}}, & x>q \\ [.08 in]
\sin (k_xx),  & -q \leq x \leq q \\ [.08 in]
\displaystyle{\frac{\cos (2k_xq) \exp [\kappa (x+q)] - \exp [-\kappa
(x+q)]}{2\sin (k_xq)}}, & x<-q. \\
\end{array}
\right.
\label{eq:thetasw}
\end{equation}

\noindent The corresponding Wronskian obeys ${\cal W}^2(\phi ,\theta ) =
2m/[\hbar ^2(ab-c^2/4)] \geq 0$ as expected.  For bound states, microstates of
the Schr\"{o}dinger wave function exist where the particular choice of the set
of coefficients $(a,b,c)$ specifies a unique trajectory in phase space for a
given quantized energy $E_x$ [\ref{bib:fpl}].  

By Eqs.\ (\ref{eq:hcfe}), (\ref{eq:eom}), (\ref{eq:phisw}) and (\ref{eq:thetasw})
and by the quantizing condition $\tan(k_xq)=\kappa /k_x$, we can evaluate the
reflection time, $t_{\pm R}$, as the time for the round trip between the barrier
interface at $x=\pm q$ and the turning point at $x=\pm \infty$.  The reflection
time, $t_{\pm R}=2|t(\pm q)-t(\pm \infty)|$, is finite for traversing an infinite
distance as given by

\begin{equation}
t_{\pm R}=2 \frac{(ab-c^2/4)^{1/2} [1+(\kappa /k_x)^2]} {a \pm c(\kappa /k_x) +
b(\kappa /k_x)^2} \frac{m}{\hbar \kappa k_x}
\label{eq:rtesw}
\end{equation}

\noindent where the sign for the coefficient $c$ in the denominator is dependent
upon which interface, $x=\pm q$, is applicable.  The trajectory for the
microstate will not be symmetric for $c\neq 0$ for our choice of $(\phi ,\theta
)$.  The existence of unsymmetric microstates of symmetric Schr\"{o}dinger wave
functions has already been discussed elsewhere [\ref{bib:prd26}].  Otherwise,
Eqs.\ (\ref{eq:rte}) and (\ref{eq:rtesw}) are very similar.  We note that
$t_{+R}$ and $t_{-R}$ are constant of the motion. 

For completeness, we present the trajectory's period for a complete libration
cycle for a bound microstate in our square well.  This libration period,
$t_{\mbox{\scriptsize{Libration}}}=2|t(\infty )-t(-\infty )|$, for a microstate
is given from Eqs.\ (\ref{eq:hcfe}), (\ref{eq:eom}), (\ref{eq:phisw}) and
(\ref{eq:thetasw}) and the quantizing condition $\tan(k_xq)=\kappa /k_x$ by

\begin{equation}
t_{\mbox{\scriptsize{Libration}}}=4 \frac{(ab-c^2/4)^{1/2} [1+(\kappa /k_x)^2]
[a+b(\kappa /k_x)^2]} {a^2 + (2ab-c^2)(\kappa /k_x)^2 + b^2(\kappa /k_x)^4}
\frac{m(q+\kappa ^{-1})}{\hbar k_x}.
\label{eq:lte}
\end{equation}  

\noindent So the libration period is a function of the coefficients $a,\ b$ and
$c$ even though, as shown elsewhere [\ref{bib:prd34}], the action variable and
energy are not.  For any allowed set of coefficients $(a,b,c)$,
$t_{\mbox{\scriptsize{Libration}}}$ is always finite.  We note that
$t_{\mbox{\scriptsize{Libration}}}$ is another constant of the motion.

In Bohmian mechanics, the one-dimensional bound state is the archetype for a
particle standing still in its original position [\ref{bib:pr85}].

We defer further comments on underdetermination until our investigation of square
well ducts.  

\section{SQUARE WELL DUCTS:}  

Let us consider a duct whose axis in two dimension ($x,y$) is aligned along the
$y$-axis.  The potential, $V(x)$, that forms the duct is still the square well
potential given by Eq.\ (\ref{eq:swv}).  The cartesian coordinate $y$ is cyclic
with a transformed constant $y$-momentum given by $\hbar k_y$ as was the case
oblique incidence given in \S \ 3.  The generator of the motion and the equation
of motion are given by Eqs.\ (\ref{eq:ocfe}) and (\ref{eq:te}) respectively where
the potential is given by Eq.\ (\ref{eq:swv}).  The Goos-H\"{a}nchen
displacement, $\Delta y_{\pm R}=2|y(\pm q)-y(\pm \infty)|$, is given by

\begin{equation}
\Delta y_{\pm R}=2 \frac{(ab-c^2/4)^{1/2} [1+(\kappa /k_x)^2]} {a \pm c(\kappa
/k_x) + b(\kappa /k_x)^2} \frac{k_y}{\kappa k_x}
\label{eq:yesw}
\end{equation}

\noindent where the sign for the subscript of $\Delta y$ and the coefficient $c$
in the denominator is dependent upon which interface, $x=\pm q$, is applicable.

The corresponding distance transversed in $y$ by the trajectory during a
libration period, $\Delta y_{\mbox{\scriptsize{Libration}}}=2|y(\infty )-y(-
\infty )|$, for a microstate is given from Eqs.\ (\ref{eq:hcfe}), (\ref{eq:eom}),
(\ref{eq:phisw}) and (\ref{eq:thetasw}) and the quantizing condition
$\tan(k_xq)=\kappa /k_x$ by

\begin{equation}
\Delta y_{\mbox{\scriptsize{Libration}}}=4 \frac{(ab-c^2/4)^{1/2} [1+(\kappa
/k_x)^2] [a+b(\kappa /k_x)^2]} {a^2 + (2ab-c^2)(\kappa /k_x)^2 + b^2(\kappa
/k_x)^4} \frac{k_y(q+\kappa ^{-1})}{k_x}.
\label{eq:ylte}
\end{equation}  

We now know how the trajectory (microstate) in a duct behaves according to its
set of coefficients ($a,b,c$).  In order to resolve underdetermination, we shall
now develop a proposed test whether an observed overdetermined set of constants
of the motion, that is accessible to the Copenhagen interpretation, is consistent
with theoretical redundancy.  

While we have been able for a given $E_x$ or $J$ to describe the set of
coefficients ($a,b,c$) in terms of the set of initial conditions
[$x_o,\dot{x}_o,\ddot{x}_o$] in the spirit of Hamilton, the Copenhagen school
would argue that we cannot measure these initial conditions simultaneously due
to the Heisenberg uncertainty principle (but we could identify either $x$ or
$\dot{x}$; note that the Heisenberg uncertainty principle does not address
$\ddot{x}$).  The Copenhagen school would conclude that we can specify neither
the coefficients $a,\ b$ and $c$ nor subsequently the microstate, which is
consistent with Copenhagen postulate that $\psi $ is an exhaustive description
of nonrelativistic quantum phenomena.  In the following three paragraphs, the set
of coefficients ($a,b,c$) shall be specified by another set of constants of the
motion that are measurable by the Copenhagen school.

We can also express the set of coefficients ($a,b,c$) in terms of constants of
the motion in the spirit of Jacobi.  (Note that the coefficients $a,\ b$ and $c$
are, in their own right, constants of the motion.)  The constants of the motion
$E_x$ and $J$ are independent of ($a,b,c$) [\ref{bib:prd34}].  Let us survey what
other constants of the motion we have that are dependent on the coefficients
($a,b,c$).  The Wronskian, ${\cal W}^2(\phi,\theta) = 2m/[\hbar ^2(ab-c^2/4)]$,
provides such a constant of the motion.   In addition, $\Delta y_+$, $\Delta y_+$
and $\Delta y_{\mbox{\scriptsize{Libration}}}$ are all such constants of the
motion.  Also, there exists a constant of the motion, $I$, which is an Ermakov
invariant established by the Ermakov system formed by the generalized 
Hamilton-Jacobi equation, Eq.\ (\ref{eq:hje}), and the Schr\"{o}dinger equation. 
This constant of the motion is given for bound states by [\ref{bib:pla214}]

\begin{eqnarray*}
I & = & \{W'\psi ^2 + (\hbar ^2/W')[\psi W''/(2W') + \psi ']^2\}/(2m)^{1/2}  \\
  & = & [a-c^2/(4b)]^{-1} = b\hbar ^2{\cal W}^2/(2m) > 0.
\end{eqnarray*}

\noindent Hence, $I$ is positive definite for bound states (it is zero for
unbound states) [\ref{bib:pla214}].

We may, for a given $E$ or $J$, describe the microstate in terms of other
constants of the motion instead of the set of coefficients $(a,b,c)$.  This would
remove the Copenhagen school's objection regarding specifying the set of
coefficients ($a,b,c$) by the set of initial conditions
[$x_o,\dot{x}_o,\ddot{x}_o$].  Three independent constants of the motion are
necessary and sufficient to specify the set of coefficients $(a,b,c)$.  We have
already developed a redundant set of five constants of the motion $(I,{\cal
W},\Delta y_+,\Delta y_-,\Delta y_{\mbox{\scriptsize{Libration}}})$ that have
been expressed herein as functions of the set of coefficients ($a,b,c$).  The set
$(b,{\cal W},I)$ is redundant as $b$ can be specified by 

\[
b=\frac{2m}{\hbar ^2}\frac{I}{{\cal W}^2}.
\]

\noindent The coefficients $a$ and $c$ are specified by

\[
a= \frac{(2m)^{1/2}}{\hbar {\cal W}^2}[1+(\kappa /k_x)^2]\frac{k_y}{\kappa k_x}
\left( \frac{1}{\Delta y_+}+\frac{1}{\Delta y_-}\right) - \frac{2mI}{\hbar {\cal
W}^2} \left (\frac{\kappa }{k_x}\right) ^2 
\]

and

\[
c=\frac{(2m)^{1/2}}{\hbar {\cal W}^2}[1+(\kappa /k_x)^2]\frac{k_y}{\kappa k_x}
\left( \frac{1}{\Delta y_+}-\frac{1}{\Delta y_-}\right).
\]

The set of Goos-H\"{a}nchen displacements ($\Delta y_+,\Delta y_-$) and $\Delta
y_{\mbox{\scriptsize{Libration}}})$ are measurable in the Copenhagen
interpretation for there is an operator for $y$ in the Schr\"{o}dinger
representation.  The Copenhagen school may still demurer with the objection that
the set $(I,{\cal W},a,c)$ is redundant and that we need one more constant of the
motion independent of this set to specify the set ($a,b,c$) by accessible
measurements.  The set of initial conditions provides us this other constant. 
The Copenhagen school must stipulate that we can know either $x_o$ or $\dot{x}_o$
by the Heisenberg uncertainty principle and perhaps even also know $\ddot{x}_o$. 
Thus, the set $(\Delta y_+,\Delta y_-,\xi )$ where $\xi $ is either $x_o$ or
$\dot{x}_o$ determines the set ($a,b,c$) in a means acceptable to the Copenhagen
school.   

We can now specify the microstate and consequently the trajectory for a particle
in a way acceptable to the Copenhagen school.  We still have the constant of the
motion, $\Delta y_{\mbox{\scriptsize{Libration}}}$, which has not been used to
specify the set of coefficients ($a,b,c$).  With the addition of $\Delta
y_{\mbox{\scriptsize{Libration}}}$, we have overdetermined the set of
coefficients ($a,b,c$).  We may now propose a test of whether the observed
overdetermination of a microstate is consistent with the theoretical redundancy
in the set of constants of the motion $(\Delta
y_{\mbox{\scriptsize{Libration}}},\Delta y_+,\Delta y_-,\xi )$ where $\xi $ is
either $x_o$ or $\dot{x}_o$.  This set is accessible to the Copenhagen school. 
Thus, this redundancy is a strong test for resolving the underdetermination
issue. If observation is consistent with the theoretical dependence among the
super-sufficient set of constants of the motion, then microstates are so, and the
Copenhagen interpretation is in variance.
 
\bigskip

\paragraph{References}
\begin{enumerate}\itemsep -.06in
\item \label{bib:rmp61} E. H. Hauge and J. A. Stovneng, {\it Rev.\ Mod.\ Phys.}\ {\bf
61}, 917 (1989).  
\item \label{bib:pla167} J. G. Muga, S. Brouard and R. Sala, {\it Phys.\ Lett.}\ {\bf
A 167}, 24 (1992).
\item \label{bib:pr214} V. S. Olkhovsky and E. Recami, {\it Phys.\ Rep.}\ {\bf 214},
339 (1992).
\item \label{bib:pla197} C. R. Leavens, {\it Phys. Rev.}\ {\bf A 197}, 84 (1995).
\item \label{bib:prd26} E. R. Floyd, {\it Phys.\ Rev.}\ {\bf D 26}, 1339 (1982).
\item \label{bib:prd29} E. R. Floyd, {\it Phys.\ Rev.}\ {\bf D 29}, 1842 (1984).
\item \label{bib:prd34} E. R. Floyd, {\it Phys.\ Rev.}\ {\bf D 34}, 3246 (1986).
\item \label{bib:fpl} E. R. Floyd, {\it Found.\ Phys.\ Lett.}\ {\bf 9},489 (1996).
\item \label{bib:jtc} J. T. Cushing, {\it Quantum Mechanics: Historical
Contingency and the Copenhagen Hegemony} (The University of Chicago Press,
Chicago, 1994) pp.\ xi--xii, 199--203.
\item \label{bib:pr85} D. Bohm, {\it Phys.\ Rev.}\ {\bf 85}, 166 (1952).
\item \label{bib:pe7} E. R. Floyd, {\it Phys.\ Essays} {\bf 7}, 135 (1994).
\item \label{bib:ap1} F. Goos and H. H\"{a}nchen, {\it Ann.\ Phys.}\ (Leipzig) {\bf 1},
33 (1947).
\item \label{bib:rmp66} R. Landauer and Th. Martin, {\it Rev.\ Mod.\ Phys.}\ {\bf 66},
217 (1994).
\item \label{bib:prl74} A. M. Steinberg, {\it Phys.\ Rev.\ Lett.}\ {\bf 74}, 2405 (1
995).
\item \label{bib:afb20} E. R. Floyd, {\it An.\ Fond.\ Louis de Broglie} {\bf 20}, 263
(1995).
\item \label{bib:pr172} D. Bohm and B. J. Hiley, {\it Phys. Rep.}\ {\bf 172}, 93 (1989).
\item \label{bib:pla214} E. R. Floyd, {\it Phys.\ Lett.}\ {\bf A 214}, 259 (1996).
\item \label{bib:ap166} G. Barton, {\it Ann.\ Phys.}\ (NY) {\bf 166}, 322 (1986).
\item \label{bib:jap33} T. E. Hartman, {\it J. App.\ Phys.}\ {\bf 33}, 3247 (1962).
\item \label{bib:jpc18} J. R. Fletcher, {\it J. Phys.}\ {\bf C 18}, L55 (1985).
\end{enumerate}

\end{document}